\documentclass[letterpaper,journal]{IEEEtran}

\usepackage{amsmath,amsfonts}
\usepackage{algorithmic}
\usepackage{array}

\usepackage{graphicx}
\usepackage[caption=false,font=normalsize,labelfont=sf,textfont=sf]{subfig}
\usepackage{xcolor}

\usepackage[T1]{fontenc}
\usepackage{textcomp}
\usepackage{url}
\usepackage{verbatim}
\usepackage{balance}

\usepackage{hyperref}
\hypersetup{
    colorlinks=true,
    linkcolor=black,
    filecolor=black,
    urlcolor=violet,
    citecolor=blue,
    frenchlinks=false
}

\usepackage{stfloats}

\def\BibTeX{{\rm B\kern-.05em{\sc i\kern-.025em b}%
    \kern-.08em T\kern-.1667em\lower.7ex\hbox{E}\kern-.125emX}}
\def\LB{\textit{LiteBIRD}}
\def\Planck{\textit{Planck}}

\begin{document}

\title{TES Proton Irradiation Result Analysis for Future Space Applications}

\author{A. Besnard$^1$, V. Sauvage$^1$, S. L. Stever$^1$, B. Maffei$^1$, 
P. dal Bo$^2$, T. Lari$^2$, M. de Lucia$^2$, A. Tartari$^2$, 
G. Signorelli$^2$, J. Hubmayr$^3$, G. Jaehnig$^3$, for the \LB\ Collaboration%
\thanks{Contact: anais.besnard@universite-paris-saclay.fr}\\
$^1$ IAS, Université Paris-Saclay, CNRS, Institut d’Astrophysique Spatiale, Orsay, France\\
$^2$ INFN, Pisa University, Istituto Nazionale di Fisica Nucleare, Italia\\
$^3$ NIST Boulder Laboratory, National Institute of Standards and Technology, Colorado, United States%
}

\maketitle

\begin{abstract}
As observed on the signal of the \Planck-HFI highly sensitive bolometers, the effect of cosmic rays on detectors is a major concern for future similar space missions. Their instruments will have a larger detection surface, increased sensitivity, and more stringent requirements on the suppression of systematic effects. To study the impact of cosmic rays on detector prototypes in operational conditions, IAS has designed a state-of-the-art cryogenic system to irradiate particles by coupling this facility to particle accelerators. An irradiation campaign has been carried out on \LB-HFT TES prototypes to study their response to particle hits. In this article, we present the results and the analysis of this first test campaign.
\end{abstract}


\section{Introduction}

\IEEEPARstart{T}{he} \Planck\ telescope is the last space mission devoted to the \textbf{C}osmic \textbf{M}icrowave \textbf{B}ackground (CMB) observation, providing the community with outstanding results. However, soon after its first observations in 2009, the \textbf{H}igh \textbf{F}requency \textbf{I}nstrument (HFI) showed that the signals from its 52 bolometers operating at 100~mK were affected by higher than expected noise. It became clear that cosmic rays (CRs), originating from outside the galaxy and with energies ranging from 0.1 to 1~GeV at the L2 Lagrangian point (where the \Planck\ satellite was located), interacted with the bolometers by depositing energy through hits, producing spurious signals appearing as glitches in the raw data. 

\begin{figure}[!ht]
\centering
\includegraphics[width=\linewidth]{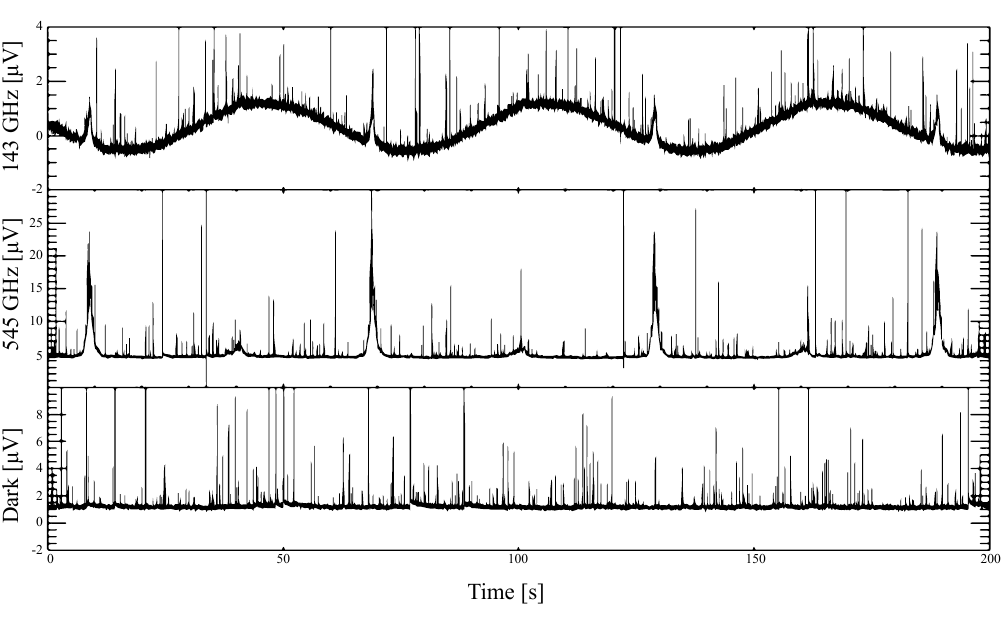}
\caption{Output signals from three bolometric detectors from the \Planck\ HFI \cite{catalano_characterization_2014}. Top: 143~GHz channel showing the CMB dipole and the Galactic center. Middle: 545~GHz channel mainly detecting the Galactic center. Bottom: blind bolometer output. Each bolometer is affected by numerous glitches.}
\label{glitches_planck}
\end{figure}

As shown in Fig.~\ref{glitches_planck}, most spikes also affected a blind detector due to CR hits. During the mission lifetime, these glitches would have resulted in about 15\% data loss, but were reduced to a few percent through post-launch analyses and complementary laboratory studies.

The impact of cosmic rays on bolometric detectors have already been studied in \cite{caserta_radiation_1990}, \cite{masi_effect_2010}, \cite{catalano_characterization_2014} and \cite{stever_new_2018}, but with future space missions planning to make use of highly sensitive detectors operating between 50~mK and 150~mK, such as \textbf{T}ransition \textbf{E}dge \textbf{S}ensors (TES) or \textbf{K}inetic \textbf{I}nductance \textbf{D}etectors (KIDs), it is crucial to evaluate the susceptibility to particle hits and to adapt their design for better immunity ahead of any mission launch, during the project development phase. For this purpose, following the CR studies initiated by the \Planck-HFI consortium, the IAS has developed a state-of-the-art cryogenic facility to irradiate cryogenic detectors, detector wafers, and focal plane components, which can be operated within a temperature range of 20 to 800~mK in a typical case. The DRACuLA (\textbf{D}etector ir\textbf{RA}diation \textbf{C}ryogenic faci\textbf{L}ity for \textbf{A}strophysics) facility can be coupled to an accelerator beam line, allowing for irradiation with particles of various energies and fluxes at the mixing chamber stage where the detector under test is placed.\footnote{The energy ranges and the particle flux depends on the accelerator installation} \\

The future \LB\  mission \cite{2023PTEP.2023d2F01L}, due to map the CMB anisotropies $B$-mode polarization across the full sky for the first time, requires a higher sensitivity than its predecessor. It is currently under development and scheduled for launch to the L2 point mid-2030s. Given that TES technology is chosen to be the detector baseline, a comprehensive study is essential to assess and quantify their sensitivity to CRs. During a test campaign in May 2024, \LB\  TES prototypes developed by the United States \textbf{N}ational \textbf{I}nstitute of \textbf{S}tandards and \textbf{T}echnology (NIST) were integrated within the DRACuLA facility and pre-tested with an internal $\alpha$ source emitting particles at 5.4 MeV. DRACuLA was finally coupled to the IJCLab ALTO particle accelerator based in Orsay, on the Paris-Saclay university campus (France) to explore higher energies. Protons with energies of 18~MeV and 22~MeV successfully irradiated these detectors at a temperature of 100~mK. 

Alongside with a brief presentation of the DRACuLA facility, we describe here the setup of this specific experiment, followed by the analysis of the results obtained during the irradiation campaign. Modeling of the glitches triggered by particle impacts on the detector allows us to predict the associated noise on future observations.

\section{Presentation of the experimental setup}

The IAS has started to study the interaction between highly sensitive cryogenic detectors and cosmic rays, right after the first data of the \Planck-HFI were received, showing a higher than expected glitch rate. These interactions can be experimentally reproduced by irradiating particles, such as protons for instance, onto operating detectors. More recently, the IAS has developed an entirely customized cryogenic facility in order to study the impact of high energy particles on new detectors and focal planes that will be part of future space missions. The first irradiation campaign with the DRACuLA facility coupled to TANDEM, a particle accelerator at the ALTO facility from IJCLab, was carried out in September 2022 on a composite semiconductor bolometer from the DIABOLO experiment \cite{desert_diabolo_2002}. The operation of the cryogenic system has been demonstrated, and with few adjustments described in \cite{besnard_cryogenic_2024}, a second campaign was carried out in May 2024, with the aim to analyse the impact of high energy particles on TES prototypes for \LB.

\subsection{Cryogenic facility} \label{cryo_fac}

DRACuLA is a dilution refrigerator, model LD400 from Bluefors, with a 500~µW of cooling power at 100~mK, hosted within a mobile compact frame. Four KF-50 ports allows for the samples to be irradiated with different incidence angles (0$^\circ$, 45$^\circ$, 90$^\circ$ and 180$^\circ$).  These particular features make it adaptable to various particle accelerator. During the \LB\  test campaign, this system has demonstrated an excellent thermal stability, with 100 mK $\pm$ 11 µK at detector level, for more informations refer to \cite{besnard_cryogenic_2024}.

\subsection{TES under test}\label{dut}

\begin{figure}[!h]
    \centering
    \includegraphics[width=3.5in]{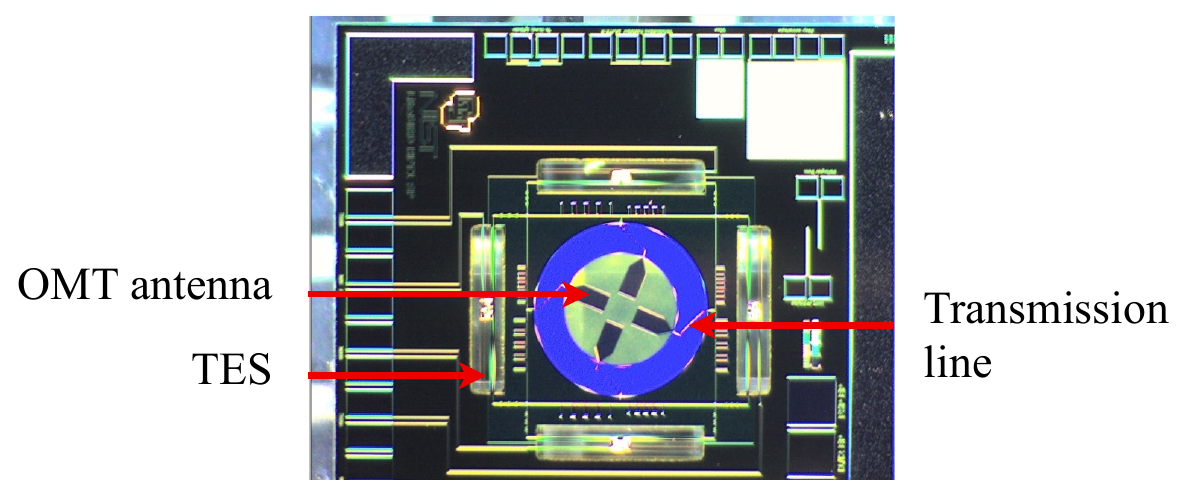}
    \caption{NIST HFT pixel prototype: a planar OMT split the incoming EM radiation into two linear polarizations before bringing the signal to filters and to the four TESs through transmission lines}
    \label{NIST_TES}
\end{figure}

The detector under test provided by NIST is a TES prototype \cite{lueker_thermal_2009}, \cite{hubmayr_optical_2022} developed for the \LB\  \textbf{H}igh \textbf{F}requency \textbf{T}elescope (HFT). This dual-polarization, dual-frequency band pixel is to be coupled to the waveguide of a feedhorn through a planar \textbf{O}rtho-\textbf{M}ode-\textbf{T}ransducer (OMT). A transmission line connects each of the four OMT probe (Fig~\ref{NIST_TES}) to a filter bank splitting the signal in two frequency bands before reaching a resistor load and the final TES.

The TES is biased around its critical temperature $T_\textrm{C}$, so that any small change in the incoming signal, dissipated into heat, will produce a large response. The TES therefore acts as a thermal detector, also sensitive to any other heat sources, such as current dissipation, and in our specific case, particle hits on any part of the pixel and surrounding structure.

\subsection{Detector test set-up}\label{prep}

\begin{figure}[!ht]
    \centering
    \includegraphics[width=3.5in]{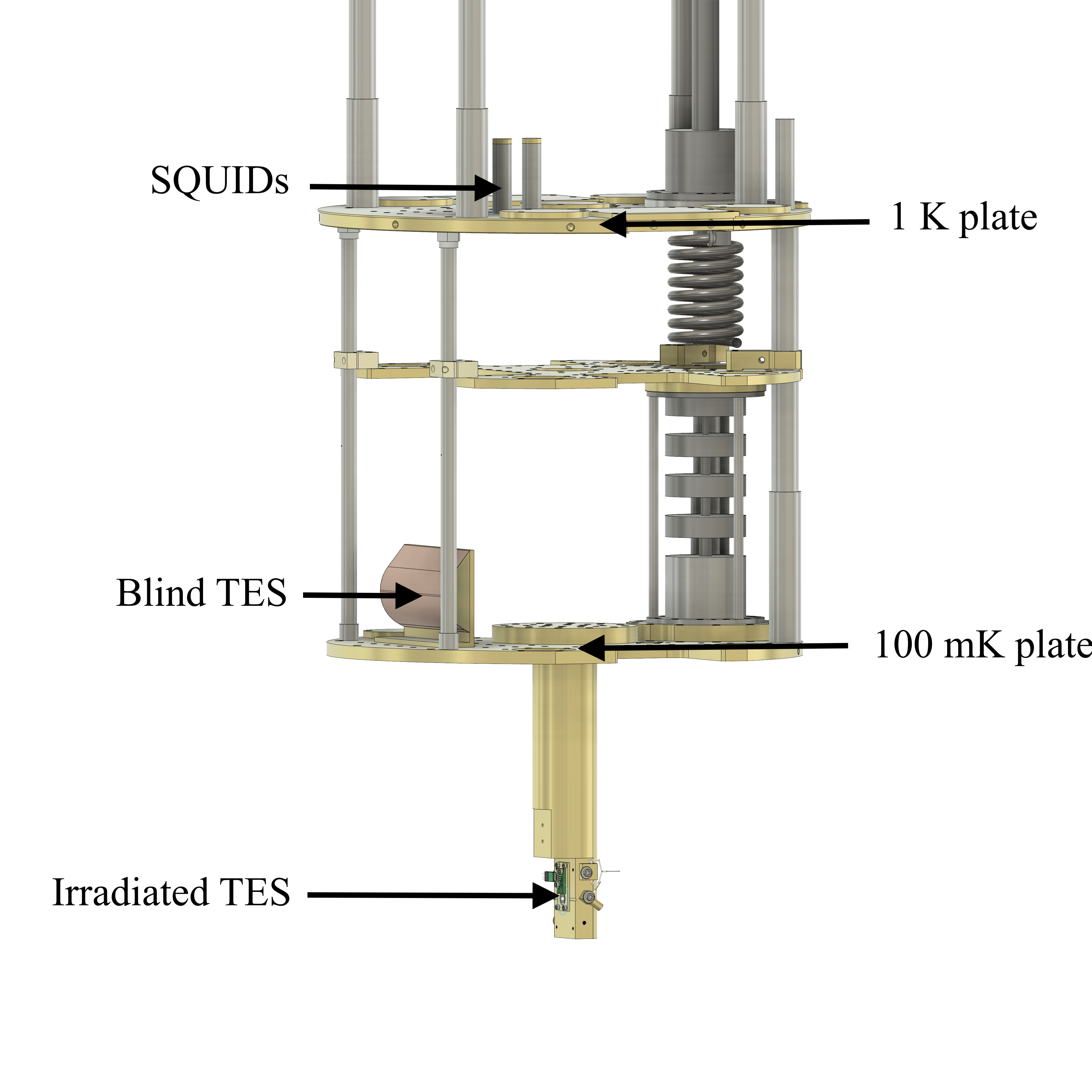}
    \caption{View of devices under test within the DRACuLA cryogenic facility.}
    \label{DRACuLA}
\end{figure}
To correlate noise contaminations from sources other than the particle hits, such as the electronics for instance,  two identical TESs were used; one was irradiated while the other was away from the beam and therefore ''blind'. Coincident events detected by both were removed  from the final dataset. 

\begin{figure}[!ht]
    \centering
    \includegraphics[width=1\linewidth]{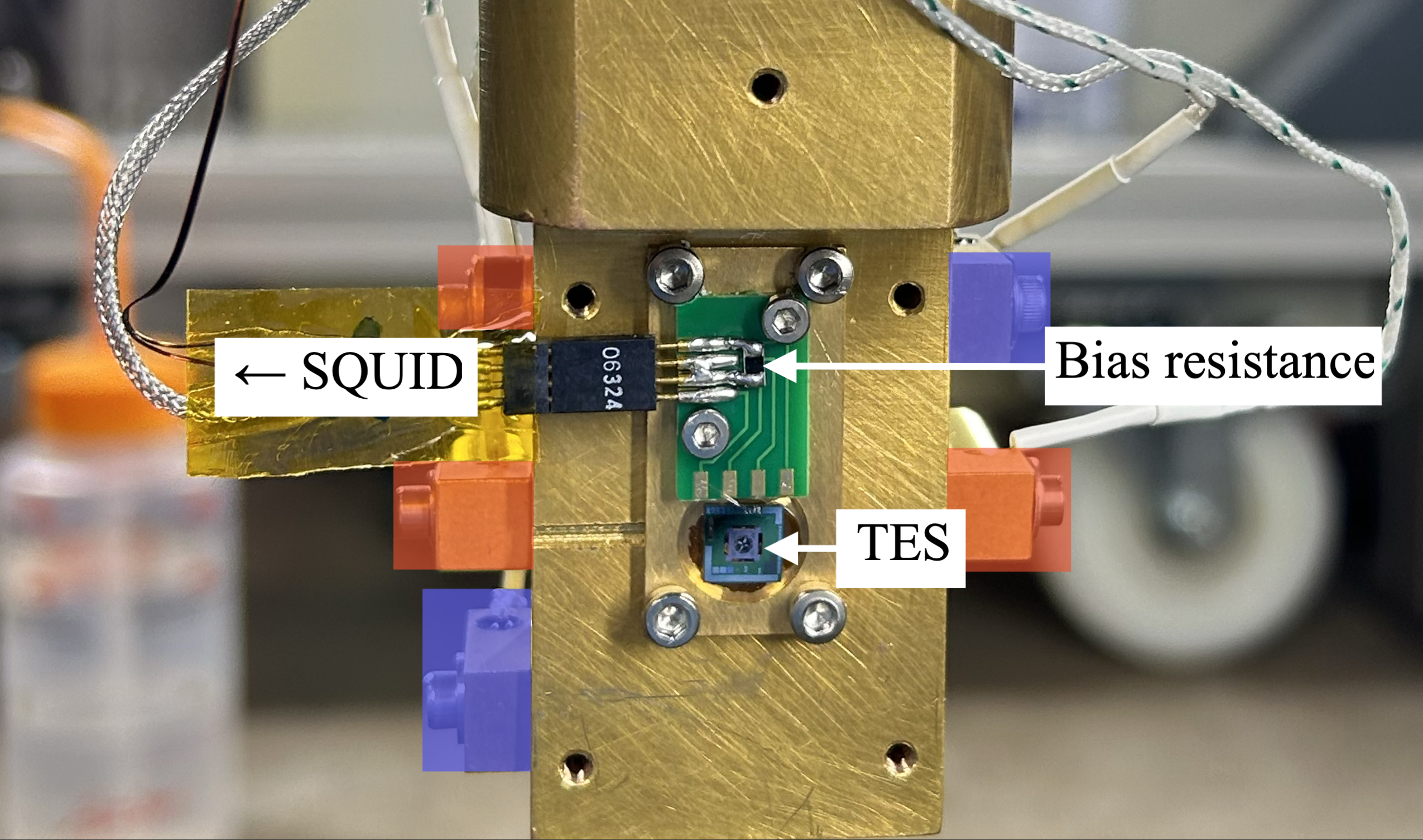}
    \caption{Close-up view of the detector under test mounted in parallel with the R$_\textrm{{bias}}$ and bounded with Al wires to the PCB, ensuring the connexion to the SQUID. The final assembly is  placed on the holder and equipped with two heaters (blue) and three thermometers (orange)}
    \label{zoom-det}
\end{figure}

The irradiated detector was placed on a holder thermalized at 100~mK in front of a KF-50 port to face the particle accelerator beamline, whereas the blind detector was installed onto the 100-mK plate with a copper shield around it. Both TESs were bounded to a custom designed \textbf{P}rinted \textbf{C}ircuit \textbf{B}oard (PCB) via 25 µm aluminum wires, through electron beam procedure at CSNSM-IJCLab in Orsay. Each PCB was connected to a \textbf{S}uperconducting \textbf{QU}antum \textbf{I}nterface \textbf{D}evice (SQUID) thermalized on the 1-K plate and used as readout. The irradiated TES was read with a StarCryo SQUID, while the blind TES was read with a VTT SQUID. Fig~\ref{DRACuLA} and \ref{zoom-det} show the setup scheme within the DRACuLA facility.
\\

The irradiated TES was also equipped with two thin metallic resistances (2 k$\Omega$ and 3 k$\Omega$) as heaters and three thermometers to control the temperature : a Cernox CX1010 for the higher temperatures and two ROx-102B for the lower temperatures and redundancy (Fig~\ref{zoom-det}). A R$_{\textrm{bias}}$ of 33-m$\Omega$ resistance was also soldered on the PCB in parallel with the TES. The bias voltage is generated by the 5-V Yokogawa 7651 output and previously reduced by a 50-k$\Omega$ resistor.

Pre-irradiation critical temperature ($T_\textrm{C}$) measurements were performed on both TESs, by simply changing the bath temperature and observing the transition, to be then compared to post-irradiation $T_\textrm{C}$ measurements, thus checking for any changes. A more precise determination of the $T_\textrm{C}$ has been also performed through measurement of current-voltage (IV) curves, and the results of these measurements are the subject of a companion paper \cite{dal_bo_2025}. \\

Prior to the move of the cryogenic facility to the particle accelerator, a first irradiation campaign at 5.4~MeV was performed with a $^{241}$Am internal $\alpha$ source at IAS. This $\alpha$ source was used in previous works \cite{stever_new_2018}, \cite{stever_characterisation_nodate} and \cite{stever_simulations_2021-1} on composite bolometers (similar to those used in \Planck) to study pulses generated by particles. This preliminary irradiation was carried out in order to ensure a good operation, but above all to obtain a first assessment of the TES susceptibility to particle hits, resulting in 149 detected events. Following this, the cryogenic facility was moved and aligned with the proton beam at the TANDEM facility \cite{vergnes_tandem_1977}. Two energies were explored: 18 MeV and 22 MeV, with a total of 2936 events observed (excluding coincident events). As expected, more saturated glitches were recorded at 5.4 MeV, due to the energy difference; particles of lower energy deposited more energy than the higher energy particles due to the effects of stopping power. 

\section{Analysis}

Following the test campaign at TANDEM, the datasets from the internal irradiation with the $^{241}$Am source at 5.4 MeV and the irradiation at 18~MeV were analyzed. Three categories of shapes were observed in both datasets, and probable hit zones on the detector chip were identified as described in section~\ref{zone&shape}. A MCMC fitting procedure has been developed to extract the time constant, a crucial parameter to model the detector that will later allow for glitch removal from the data. This procedure is detailed in section~\ref{MCMCfitting}, together with an additional FFT approach to validate or improve the pulse fit.

\subsection{Relationship between pulse shape and heat deposition location} \label{zone&shape}

\begin{figure}[!h]
    \centering  \includegraphics[width=1\linewidth]{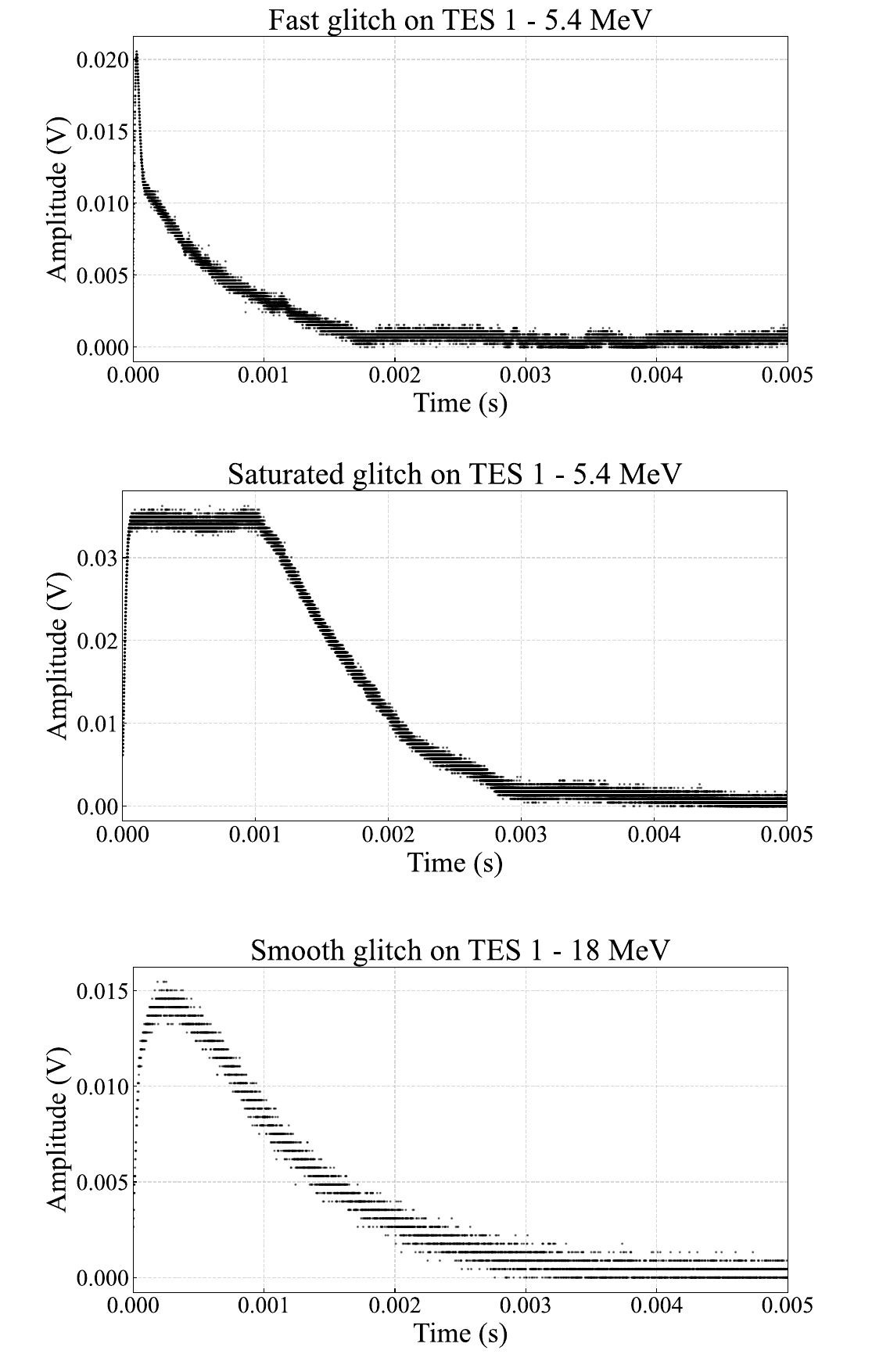}
    \caption{The different shapes identified in the data as (a) fast glitch, (b) saturated glitch, (c) smooth glitch}
    \label{shapes}
\end{figure}

\begin{figure}[!ht]
    \centering
    \includegraphics[width=0.85\linewidth]{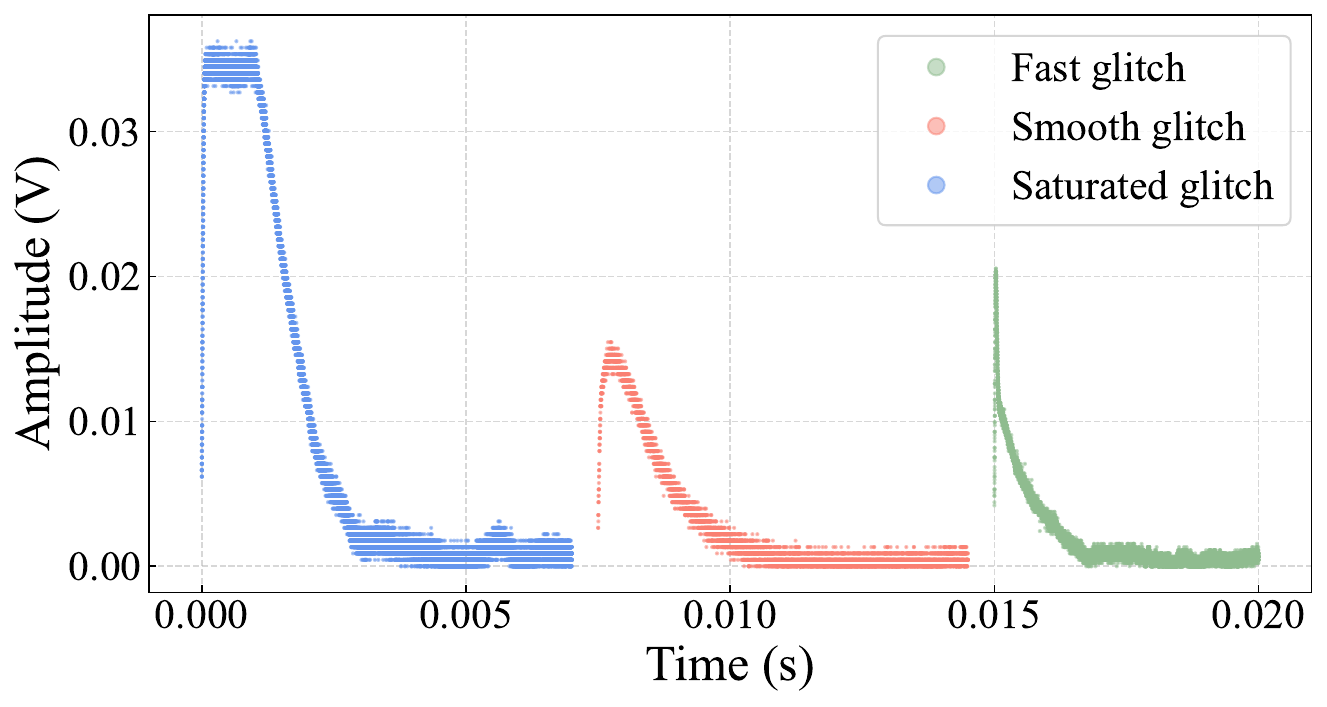}
    \caption{Gitch shape comparison}
    \label{shape_comparison}
\end{figure}

A first analysis of the $\alpha$ particles (5.4~MeV) and protons (18~MeV) led to three main categories of pulse profiles. Each shape is presented individually over the same time range in Fig~\ref{shapes} and together in Fig~\ref{shape_comparison} (for comparison).  \\

% \begin{figure}[!ht]
%     \centering
%     \begin{subfigure}
%         \includegraphics[width=1\linewidth]{0_Figures/fast_54.pdf}
%         \label{fast}
%     \end{subfigure}
%     \begin{subfigure}
%         \includegraphics[width=1\linewidth]{0_Figures/sat_54.pdf}
%         \label{sat}
%     \end{subfigure}
%     \begin{subfigure}
%         \includegraphics[width=1\linewidth]{0_Figures/smooth_18.pdf}
%         \label{smooth}
%     \end{subfigure}

%     \caption{The different shapes identified in the data as (a) fast glitch, (b) saturated glitch, (c) smooth glitch}
%     \label{shapes}
% \end{figure}

\begin{figure}[!ht]
   \centering
   \includegraphics[width=3.5in]{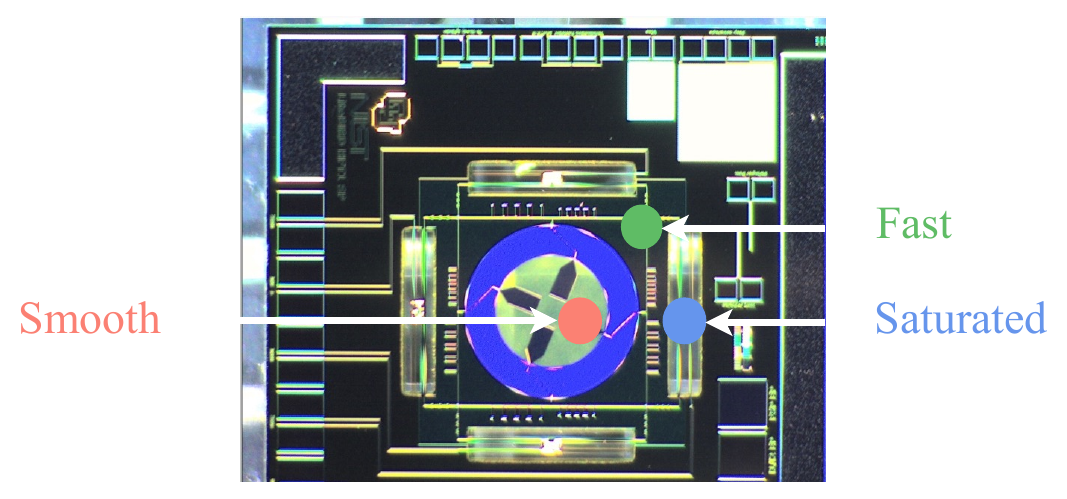}
   \caption{Representation of the relation between the pulse shape and the zone impacted by the particles. This figure is for illustrative purpose, the spots are not the precise location of the particles hits the precise location of the impacts at the origin of the glitches presented Fig~\ref{shape_comparison}}
   \label{hit zones}
\end{figure}

From this preliminary observation, we hypothesize on the topic of the relation between the shape versus the physical area absorbing the particle flux. We assumed the ``fast glitch" shown in Fig~\ref{shapes}(a) occurred on the chip between the antenna probes and the TES, shown as the green spot on Fig~\ref{hit zones}, which produces rapid rise and first decay before the exponential decrease. This `spike' is characteristic of a fast energy propogation from non-scattering `ballistic' phonons propagating to the TES, and the exponential decay results from the diffusion of normal phonons, i.e. heat. The "saturated glitch", Fig~\ref{shapes}(b) must result from a direct hit on a TES, shown as the blue spot on Fig~\ref{hit zones}. Finally, the "smooth glitch" Fig~\ref{shapes}(c) indicates a hit on the OMT, shown in red on Fig~\ref{hit zones}. This profile is driven naturally by the heat induced by a particle hit on one of the probe, which is then conducted the TES through the transmission line.

\subsection{Pulse shape fitting procedure} \label{MCMCfitting}

There are two contributions to the shape of a glitch to take into consideration : thermal and athermal propagation profiles. The glitch shape is mostly dominated by the thermal time constant of the detector, defined as $\tau_{\textrm{th}}$, resulting from the heat diffusion after rapid thermalization of the sensor caused by a particle hit. As for the athermal contribution, it originates from the ballistic phonons transporting most of the energy from the particles to the sensor at the speed of sound in the medium. This contribution is represented as a spike before the exponential decay, such as in Fig~\ref{shapes}(a) with the fast glitch. This category of glitch is known to have a faster athermal time constant $\tau_{\textrm{ath}}$ and a longer $\tau_{\textrm{th}}$, as presented in \cite{stever_characterisation_nodate}. In fact, the amplitude of the spike depends on the localization of the particle hit: the larger the spike, the closer to the detector the particle hit. There is also a correlation between the amplitude and the thermal time constant: the higher the amplitude of the phonon spike is, smaller the thermal tail will be. Thus, the time constant of a glitch is an important parameter to identify in order to properly remove the event from the data, the aim of this work. \\
\\
A first estimate of the time constant $\tau_\textrm{th}$ was done through a classic decreasing exponential function with the \texttt{scipy.optimize.curve\_fit} package from python on the pulse decay for each glitch category. The results are presented in Tab~\ref{taus_rough}, and will be compared to the MCMC procedure described below. As expected, the time constants are very similar to each other for the saturated and smooth glitch, at the different energies. Concerning the fast time constant, in both cases presented, it is longer than the others as previously predicted : the $\tau_{\textrm{th}}$ is depending on the amplitude of the spike caused by the phonon contribution, and as the amplitude is reasonably small, the time constant appear longer.

\begin{table}[!h]
\centering
\caption{Estimate of the time constants}
\label{taus_rough}
 \begin{tabular}{ccc}
 \noalign{\vskip 3pt\hrule \vskip 1.5pt \hrule \vskip 5pt}
 Shape& 5.4 MeV & 18 MeV \\ [0.5ex] 
 \noalign{\vskip 5pt\hrule\vskip 5pt}
 Smooth & 510 $\mu s$ & 453 $\mu s $\\ 
 Saturated & 500 $\mu s $& 480 $\mu s $\\ 
 Fast & 580 $\mu s$ & 520 $\mu s$ \\
 \noalign{\vskip 3pt\hrule}

 \end{tabular}

\end{table}

\begin{figure}[!ht]
    \centering
    \includegraphics[width=\linewidth]{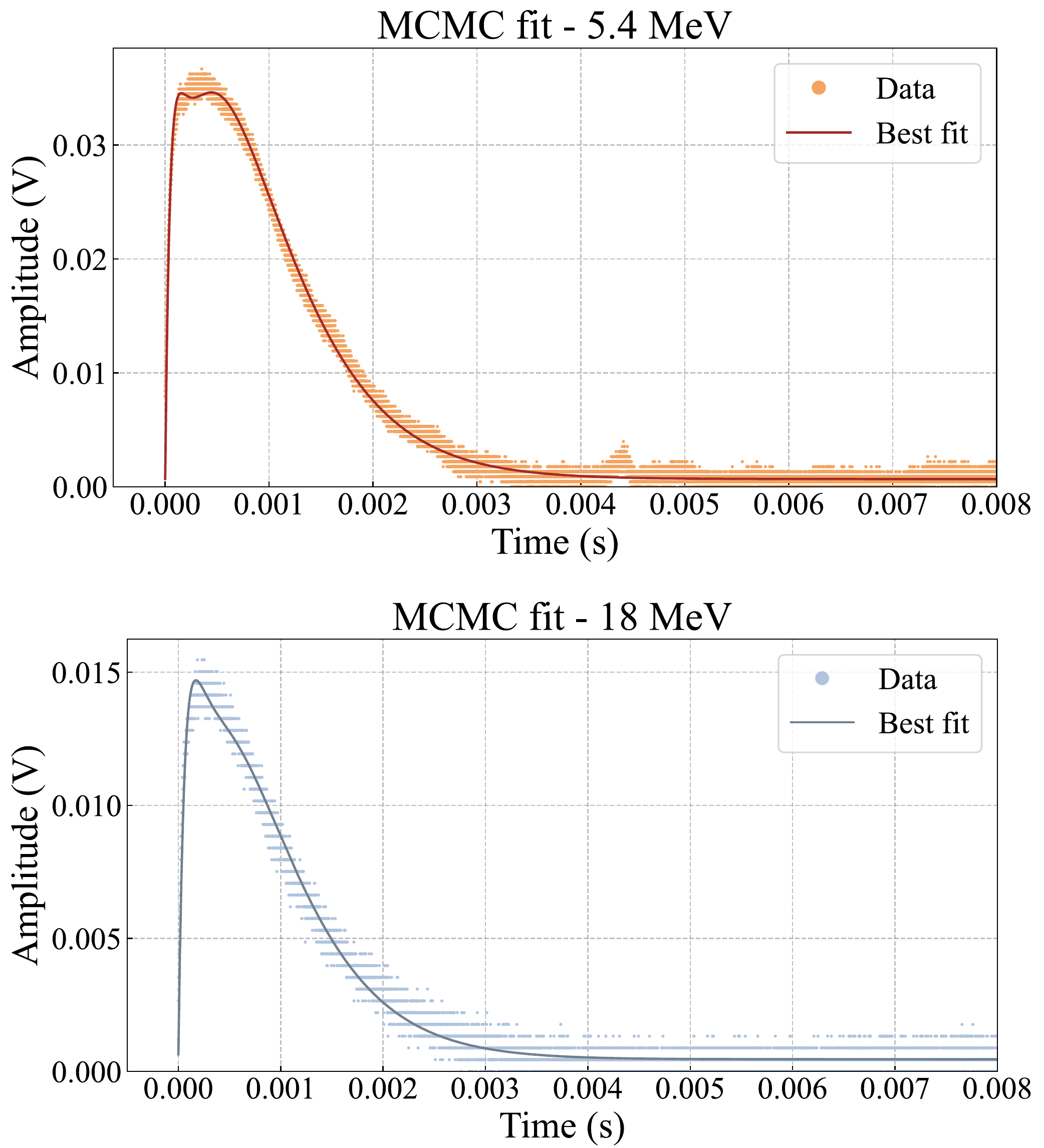}
    \caption{Fits on the "smooth glitch" for the two energies.}
    \label{fig:fit}
\end{figure}

Then, we proceeded to a more precise fit of the full shape of the glitch with a MCMC procedure. We present this method on the "smooth glitch" considering the athermal contribution and the thermal one. As in real operation, the detector chip should be shielded (mechanical structure and feedhorn) with only the antenna facing the sky and, assuming  for this preliminary work that the shield is stopping the cosmic rays without generating secondary particles on the chip, the authors decided to focus on the smooth glitch. Using the equation \ref{fit}, adapted from a double pulse fitting procedure developped for \textit{ATHENA X-IFU}'s TESs prototypes in \cite{den_herder_cryogenic_2016}, we obtain the data fit as presented in Fig~\ref{fig:fit} for the 5.4 MeV in orange and the 18 MeV in blue.\\
\begin{equation}
    P(t) = A_\textrm{ath}\frac{t}{\tau_\textrm{ath}}e^{1 - \frac{t}{\tau_\textrm{ath}}} +  A_\textrm{th}\frac{t}{\tau_\textrm{th}}e^{1 - \frac{t}{\tau_\textrm{th}}}
    \label{fit}
\end{equation}

With $A_\textrm{ath}$ the amplitude and $\tau_{\textrm{ath}}$ the time constant of the athermal phonon contribution, $A_\textrm{th}$ the amplitude and $\tau_{\textrm{th}} $ the time constant of the thermal diffusion. \\

% \begin{figure}[!ht]
%     \centering
%         \begin{subfig}
%         \centering
%         \includegraphics[width=\linewidth]{0_Figures/MCMC_fit_alpha_custom.pdf}
%         \end{subfig}
%     \hfill
%         \begin{subfig}
%         \centering
%         \includegraphics[width=\linewidth]{0_Figures/MCMC_fit_proton_custom.pdf}
%         \end{subfig}
%     \caption{Fits on the "smooth glitch" for the two energies.}
%     \label{fig:fit}
% \end{figure}

\noindent From this we extracted the time constants from the two contributions, as presented in Tab~\ref{taus}. 

\begin{table}[!hb]
\caption{Time constants of the athermal and the thermal contribution to the smooth glitch depending on the energy considered}
\label{taus}
\centering
 \begin{tabular}{ccc}
 \noalign{\vskip 3pt\hrule \vskip 1.5pt \hrule \vskip 5pt}
   & 5.4 MeV & 18 MeV \\[0.5ex]
 \noalign{\vskip 5pt\hrule\vskip 5pt}
 $\tau_\textrm{ath}$& 75.30 $\mu s ^{+0.02 \mu s}_{-0.02 \mu s}$& 95.09 $\mu s ^{+0.20 \mu s}_{-0.20 \mu s}$\\ 
 $\tau_\textrm{th}$& 505.90 $\mu s ^{+0.03 \mu s}_{-0.03 \mu s}$& 485.76 $\mu s ^{+0.30 \mu s}_{-0.30 \mu s}$\\ 
\noalign{\vskip 3pt\hrule}
 \end{tabular}

\end{table}
These results confirm that the glitch is, in fact, dominated by the thermal contribution with a $\tau_\textrm{th} = 6.7\tau_\textrm{ath}$ for the $\alpha$ irradiation at 5.4 MeV and $\tau_\textrm{th} = 5.1\tau_\textrm{ath}$ for the proton irradiation at 18 MeV.  Moreover, the thermal time constant is very close to our first estimate and the same order of magnitude at the two energies : $\tau_\textrm{th}\simeq$ 500~µs at 5.4 MeV and $\tau_\textrm{th} \simeq$ 480~µs at 18 MeV. Hence, our MCMC fitting method provide a good estimate of the thermal time constant. \\

\begin{figure}[!h]
    \centering
    \includegraphics[width=1\linewidth]{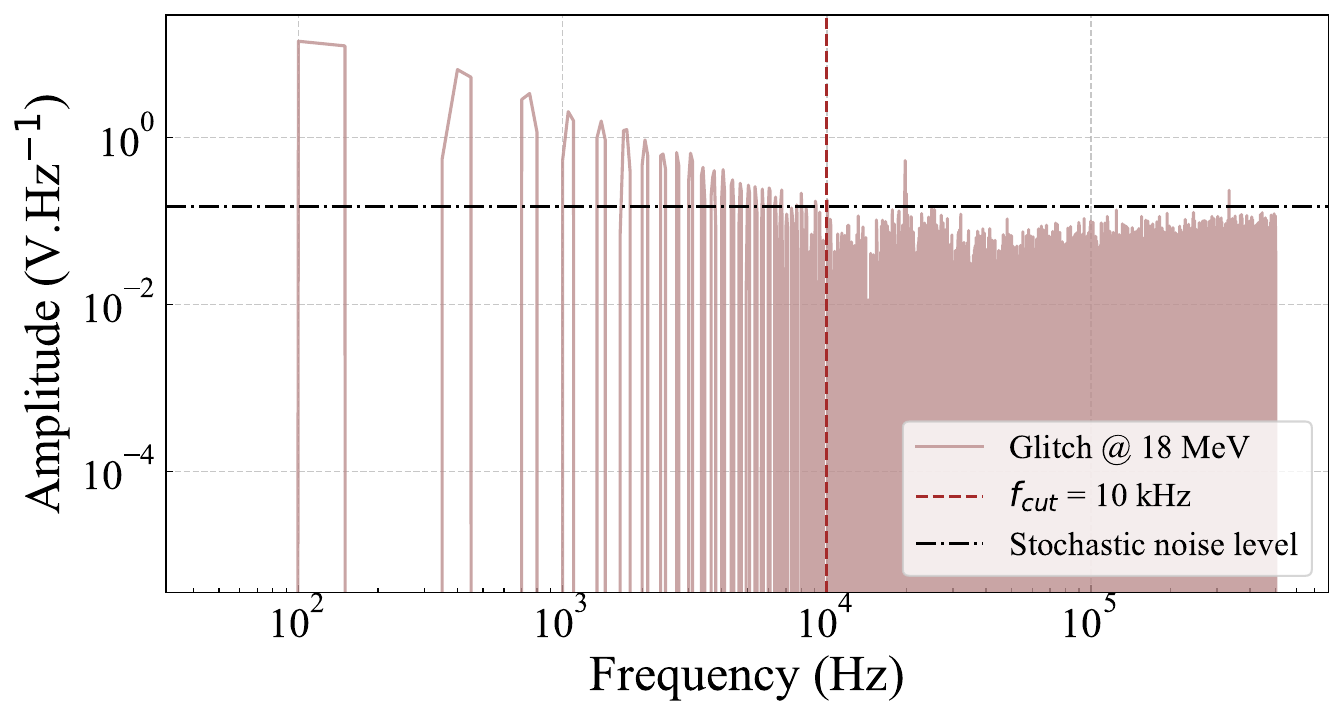}
    \caption{FFT of the smooth glitch at 18 MeV with the cut frequency at 10 kHz}
    \label{fft18}
\end{figure}

After extracting the different parameters of the fit, we subtracted the fit from the data and the recovering of the initial signal was not as perfect as expected.
Because of the acquisition noise, it is necessary to verify the accuracy of the fit in order to allow for an optimized removal of the glitch from the data. To do so, we have studied the FFT of the signal, as presented Fig~\ref{fft18}. From the FFT, we can isolate the contribution of the glitch by filtering the frequencies higher than 10 kHz, where the FFT has smaller amplitude variations and where the stochastic noise dominates.
By using an inverse FFT to recover the signal of the glitch without noise, we then fitted the isolated glitch with the MCMC procedure described above. To improve the fitting, a temporal offset was added to the time of the thermal contribution, considering the difference of speed propagation of both considered physical phenomena. The offset is determined within the MCMC procedure and is $\Delta t =$ 99~µs. 

\begin{figure}[!ht]
    \centering
    \includegraphics[width=1\linewidth]{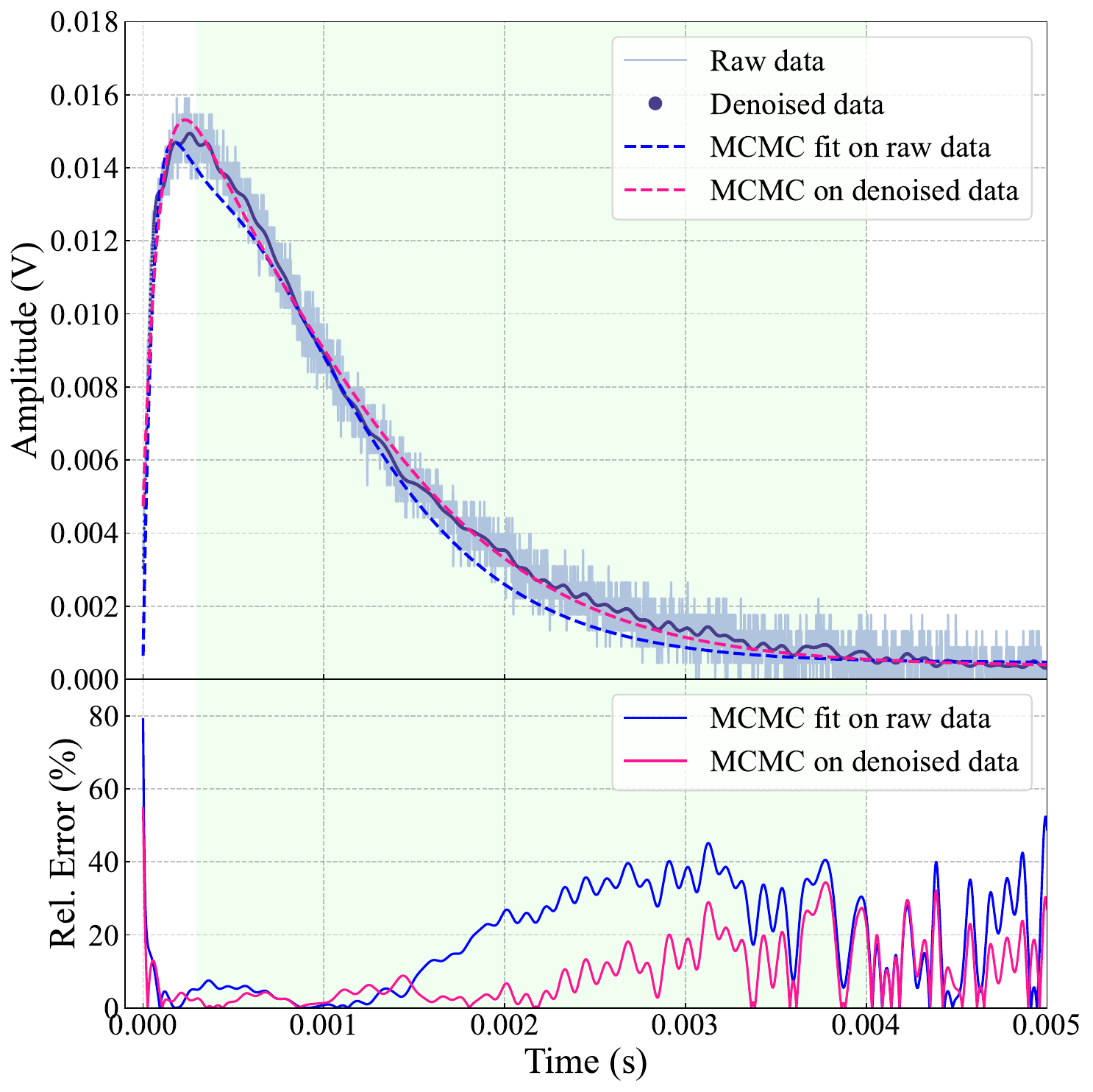}
    \caption{Comparison between the MCMC fitting procedure on the raw data and on the previously denoised data from proton irradiation at 18~MeV, with their relative errors}
    \label{new_fit}
\end{figure} 

In Fig~\ref{new_fit}, there is a comparison between the fit presented in Fig~\ref{fig:fit} and the one performed on a denoised glitch, with their relative errors. The green zone highlights where the thermal contribution dominates, and by extension, where it is important to reduce the relative error in order to improve the estimate of $\tau_\textrm{th}$.

Introducing a FFT de-noising approach and a temporal offset to the MCMC, have considerably improved the fit, allowing the authors to extract a more accurate $\tau_\textrm{th} = $ 584~µs~$^{+0.35~\mu s}_{-0.32~\mu s}$ for the smooth glitch which resulted from the proton irradiation at 18~MeV. 

\section*{Conclusion}

In this paper, we have presented the results of the first irradiation campaign on a TES prototype in preparation of the \LB\  mission. By using the unique cryogenic facility DRACuLA, a few particle energies were explored and this work allowed the authors to confirm the TES's susceptibility to cosmic rays and to obtain very encouraging results for this explorative work. By observing the different glitches recorded, we have been able to link the shape and the zone hit on the detector chip. After a rough estimate of the thermal time constant on the three shape categories, we have developed a MCMC fitting method by assuming the detector chip is in the ideal operation conditions with only the OMT antenna exposed to the cosmic rays. We have demonstrated the extraction of the thermal time constant, a critical parameter to ease the cleaning process of the data, by denoising and isolating the signal of the glitch through FFT, we have shown a method to improve the MCMC pulse fitting procedure, and by extension increased the precision of the $\tau_\textrm{th} = $ 584~µs~$^{+~0.35~\mu s}_{-~0.32~\mu s}$ of the proton irradiation at 18 MeV. 

The next step is to automatize the process of the thermal constant extraction through the MCMC fitting procedure, by using machine learning for instance. A thermal simulation of HFT, similar to what has already been done for the \LB\ 's Low Frequency Instrument \cite{stever_simulations_2021}, will be performed. Later on, characterization of the effect on the signal due to cosmic ray hits on the whole focal plane will also be checked, alongside other irradiation campaigns. 

\section*{Acknowledgments}
This work is co-funded by the French National Space Agency (CNES), the Ile de France region through the DIM-ACAV research program, the Paris-Saclay University, and the OSUPS (Observatoire des Sciences de l’Univers Paris-Saclay). The authors also thank the ALTO staff from IJCLab. Finally, special thanks to F.~Couchot for guidance on particle interactions with matter. 

\textit{LiteBIRD} (phase A) activities are supported by: ISAS/JAXA, MEXT, JSPS, KEK (Japan); CSA (Canada); CNES, CNRS, CEA (France); DFG (Germany); ASI, INFN, INAF (Italy); RCN (Norway); MCIN/AEI, CDTI (Spain); SNSA, SRC (Sweden); UKSA (UK); and NASA, DOE (USA).

% --------------------------------------------------
% BIBLIOGRAPHY
% --------------------------------------------------
\bibliography{detectors2025}

\end{document}